\documentclass[12pt,preprint]{aastex}
\shorttitle{Linear Accelerating Superluminal Motion Model}
\shortauthors{Zhou et al.}

\begin{document}

%% LaTeX will automatically break titles if they run longer than
%% one line. However, you may use \\ to force a line break if
%% you desire.

\title{Linear Accelerating Superluminal Motion Model}
\author{J. F. Zhou\altaffilmark{1}, C. Zheng\altaffilmark{2},
 T. P. Li\altaffilmark{1}, Y. Su\altaffilmark{3}, and T. Venturi\altaffilmark{4}}
\altaffiltext{1}{Department of Engineering Physics and Center for Astrophysics, Tsinghua University, Beijing 100084,China}
\altaffiltext{2}{KIPAC, Stanford University, Standford, CA 94309, U.S.A.}
\altaffiltext{3}{National Astronomical Observatories, Chinese Academy of Sciences,
    Chaoyang District, Datun Road 20A, Beijing, 100012, China}
\altaffiltext{4}{Istituto di Radioastronomia del CNR, Via Gobetti 101, I-40129 Bologna, Italy}

\begin{abstract}
Accelerating superluminal motions were detected recently by multi-epoch
Very Long Baseline Interferometry (VLBI) observations.
Here, a Linear Accelerating Superluminal Motion (LASM) model is proposed to
interpret the observed phenomena. The model provides a direct and accurate way
to estimate the viewing angle of a relativistic jet. It also predicts that both
Doppler boosting and deboosting effects may take place in an accelerating forward 
jet.
The LASM model is applied to the data of the quasar 3C 273, and the initial velocity,
acceleration and viewing angle of its three components are derived through model
fits. The variations of the viewing angle suggest that a supermassive black hole
binary system may exist in the center of 3C273. The gap between the inner and outer
jet in some radio loud AGNs my be explained in terms of Doppler deboosting effects
when the components accelerate to ultra-relativistic speed.
\end{abstract}


\section{Introduction}
Superluminal motions were first detected by Very Long Baseline Interferometry (VLBI)
in 1970s
\citep{coh77}. The observations carried out since then showed that such motions are
quite common in blazars. Many models were proposed to explain the phenomenon,
which were reviewed by \citet{bla77}. The relativistic jet model \citep{bla79} has become
the {\it de facto} standard model to describe the superluminal motion.

From a simple physical consideration, acceleration is required to
drive the subrelativistic material in the accretion disk up to the
relativistic speed observed in the jets. Acceleration is also
necessary to compensate the radiation losses in jets. There is
increasing evidence that the central engine of acceleration in
AGNs is a rotating supermassive black hole surrounded by a
geometrically thin accretion disk, which gives rise to the
formation of a pair of relativistic jets. The mechanism of
acceleration may be due to the centrifugal and shear effects
\citep{web94,rie02}. In recent years, multi-epoch VLBI monitoring
indeed led to the discovery parsec-scale accelerating superluminal motion in
the quasar 3C 273 \citep{kri01} and in other sources
\citep{hom01}.

Currently, quadratic or cubic polynomials are usually employed to fit the
accelerating superluminal motion curves \citep{kri01}. Such mathematical fits,
however, have no real physical meaning. Actually, a physical model is required to
describe the observed phenomena. Here, we consider a simple model where
a component is ejected out collinearly, with intrinsic constant acceleration
$g$. Such acceleration may exist in rotating magnetized jets \citep{rie02}, especially
at the outsets of jets where the resistance of the intergalactic medium (IGM) can be
neglected. Also, some theoretical work shows that the jets in pulsars indeed move
outward with linear acceleration \citep{con02}. In the frame of special relativity,
this model can interpret the accelerating superluminal motions, and we name it as
a Linear Accelerating Superluminal Motion (LASM) model.

In this Letter, the LASM model will be introduced in detail.
The basic kinematic equations of an
accelerating relativistic component and their solutions are listed in Section 2.
In Section 3, the proper motion of 3C\,273 monitored with VLBI is model-fitted,
and the viewing angle, initial velocity and acceleration of three components
are estimated. Possible application of the LASM model, with emphasis on the
estimates of the viewing angle and the evolution of the Doppler factor, are discussed
in Section 4.

Throughout this Letter, we adopt a flat, accelerating cosmology with
Hubble constant
$H_0 = 71 {\rm km\, s^{-1}\, Mpc^{-1}}$, $\Omega_\Lambda=0.73$ and
$\Omega_{\rm m} = 0.27$.

\section{Linear Accelerating Kinematics}
\subsection{Basic Equations \label{secbe}}
We follow the relativistic jet model \citep{bla79}. In this model, a component
moves out along a collimated jet forming a viewing angle $\theta$ with
the line of sight.
The kinematics of this component can be described by the following
equations
\begin{eqnarray}
d\mu_{\rm ob} &=& \sin\theta \,\,dr \label{basic1} \\
\beta &=& \frac{dr}{dt}  \label{basic2} \\
\frac{d\gamma\beta}{dt} &=& g \label{basic3} \\
\frac{d t_{\rm ob}}{dt} &=& 1 - \beta\cos\theta \label{basic4} \\
\beta_{\rm app} &=& \frac{d\mu_{\rm ob}}{d t_{\rm ob}},
\end{eqnarray}
where $r$ is the displacement between the component and its origin,
$\mu_{\rm ob}$ is the observed proper motion,
$t$ and $t_{\rm ob}$ are the intrinsic time and the observer's time respectively ,
$\beta$ and $\gamma=1/\sqrt{1-\beta^2}$ denote the component's velocity (in
units of the speed of light) and the Lorenz factor, $g$ is the acceleration,
and $\beta_{\rm app}$ is the apparent velocity.

In order to interpret the accelerating superluminal motions observed by VLBI, the
relationship between $\mu_{\rm ob}$ and $t_{\rm ob}$ should be derived based on the
above equations. In case of linear acceleration where $\theta$ remains constant,
the solution can be deduced through the solving routines of normal differential equations.
It is worth noting that parametric equations, which describe
the relationship between
$\mu_{\rm ob}$ and $t_{\rm ob}$, are simpler to be derived than the ``direct''
solution of $\mu_{\rm ob}$ in terms of $t_{\rm ob}$.

\subsection{Solutions \label{sec22}}
Assuming that $g$ is constant, parametric solutions of
$\mu_{\rm ob}$ and $t_{\rm ob}$ can be derived. They are
\begin{eqnarray}
\mu_{\rm ob}&=&\frac{\sin\theta}{k}\left[ \sqrt{k^2t^2+2k\beta_0 t + 1}-1 \right]
              \label{sol1}\\
t_{\rm ob} &=& t - \frac{\cos\theta}{k}\left[
\sqrt{k^2t^2+2k\beta_0 t + 1}-1 \right], \label{sol2}
\end{eqnarray}
where $k = g\sqrt{1-\beta_0^2}$, and $\beta_0$ is the initial
velocity.

$\mu_{\rm ob}$ can also be written in terms of $t_{\rm ob}$ directly.
If we eliminate $t$ in Equations \ref{sol1} and \ref{sol2}, we obtain
\begin{equation}
\mu_{\rm ob} = \frac{1}{k\sin\theta}(\sqrt{T_{1}^{2}+T_2}-T_1),
\end{equation}
where
\begin{eqnarray*}
T_1 &=& 1 - t_{\rm ob} k \cos\theta - \beta_0\cos\theta \\
T_2 &=& (t_{\rm ob}^2 k^2 + 2 t_{\rm ob} k \beta_0)\sin^2\theta .
\end{eqnarray*}

The description of the other parameters of the jet components, such as apparent velocity
$\beta_{\rm app}$ and Doppler boosting factor $D=\gamma^{-1}(1-\beta\cos\theta)^{-1}$,
can also be derived from Equations \ref{sol1}, \ref{sol2} and from the basic equations in
Section \ref{secbe}.
The corresponding formulas are
\begin{eqnarray}
\beta_{\rm app} &=& \frac{\sin\theta(k t + \beta_0)}
 {\sqrt{k^2 t^2 + 2 k \beta_0 t + 1}-\cos\theta(k t + \beta_0)} \\
D &=& \frac{ \sqrt{1-\beta_0^2}}{\sqrt{k^2t^2+2k\beta_0 t+1}-
\cos\theta(k t +
\beta_0)}. \label{doppler}
\end{eqnarray}

Compared to the non-accelerating relativistic jet models, the accelerating
model has some new interesting features.
When $t$ is infinitesimal, the initial apparent velocity will be
$\beta_{\rm ob}^0 = \beta_0\sin\theta/(1-\beta_0\cos\theta)$. As
$t$ increases, the apparent velocity will approach to its
maximum value
$\beta_{\rm ob}^{\rm max} = \cot(\theta/2)$ \citep{bla79}.
Concerning the Doppler factor $D$, it will first increase with $t$, and
it will reach its maximum value $D^{\rm max}$, where $\beta=\cos\theta$ and
$\gamma = D$. If the acceleration goes on, then $D$ will decrease, to
values even lower than 1. Therefore, in an accelerating relativistic jet, both
Doppler boosting and deboosting phenomena can be observed.

The LASM model can be applied to estimate the parameters of a relativistic jet.
Equations \ref{sol1} and \ref{sol2} show
that the shape of a proper motion curve is controlled by three
parameters, i.e. $\beta_0$, $g$, and the viewing angle $\theta$.
By comparing the model to the observed VLBI data, these parameters can be
calculated. If multi frequency flux density data were combined with proper
motion data, then the model fitting results would be more accurate.

Since the relativistic jets in AGNs probably exhibit different
structures on different scales, we should carefully consider where the 
LASM model can be applied. On the large kpc scale, the jets usually
have collinear structures. Although superluminal motions were
detected by HST observations on such scale in M87 \citep{bir99},
there are not enough data for a quantitive estimate of the jet parameters
with LASM model. However, in this situation, the LASM
model can still be used to qualitatively explain some interesting
profiles of the jets (see Section \ref{sec42}). On the scale of tens of 
pc, the jets probably show 'wiggles' or helical structures, and
the LASM model cannot be used there. On the small scale, especially
in the core regions of jets, where jet components usually move out
collinearly, the LASM is applicable to assess the physical
and geometrical parameters of jets.

\section{Application to 3C273}
3C\,273 (J1229+0203) is a low optical polarization quasar (LPQ) with $z = 0.158$
\citep{str92}. It was the first object to display apparent superluminal
motion on parsec scales \citep{pea81}. The radio structure of 3C\,273 shows a
well defined core-jet morphology from the mas scale up to the arc-second scale.
The jet extends out to ~20 arc-seconds, with the ridge line of emission showing
a clear 'wiggle' \citep{dav85}. VLBI imaging at frequencies from 5 to 100 GHz
show that the ridge line of the jet is curved on a scale from 0.05 to
25 mas, oscillating around the main orientation of the
jet \citep{baa91,kri90}. \citet{zen90} suggest that the discrepancy in the position
angles of some of the
components observed at different frequencies may reflect a spectral index
gradient across the jet.

Since 1990, 3C\,273 has been monitored with VLBI from 15 to 86 GHz.
For the components with enough data points at small ($<$ 2mas) and large
($>$ 2mas) core separations, quadratic fits to the radial motion $r(t)$
describe the observations much better than linear fits \citep{kri01}. This 
suggests that these components indeed have accelerating proper motions.
%indicates
%that these components indeed have accelerating proper motions.
Furthermore, in the inner region with radial distance less than 8 milli-arcsecond,
these components seem to move outward collinearly.
Therefore, the LASM model is suitable to describe
the observed data, and estimate the  viewing angles and accelerations
the components via least square fits.

\begin{figure}
\includegraphics[width=10cm]{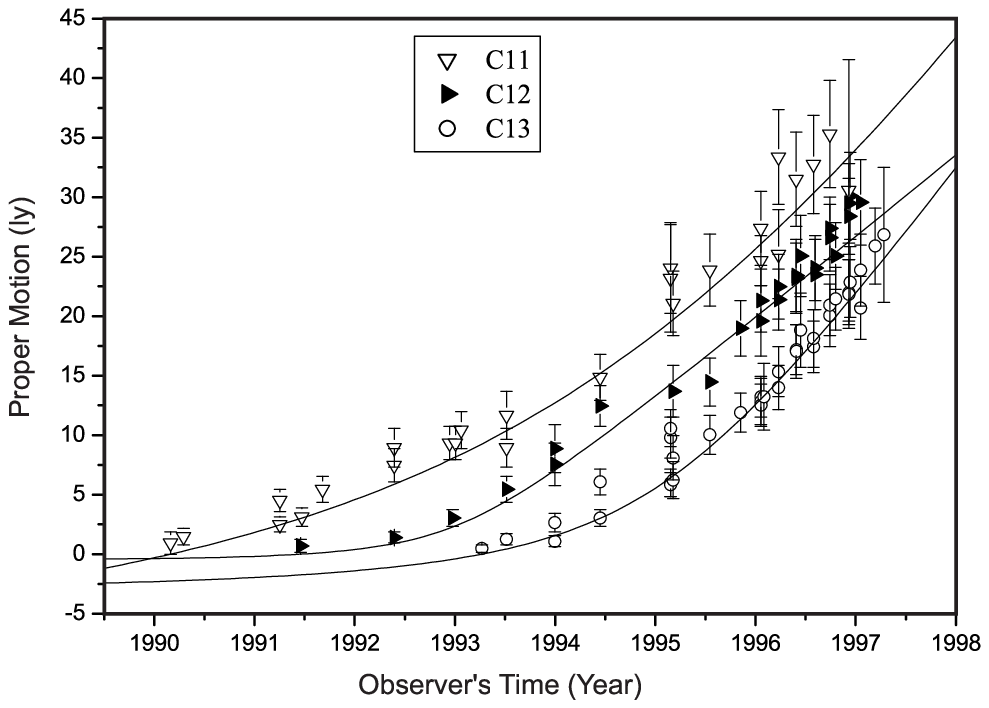}
\figcaption[f2.eps]{LASM model fitting results of the components C11, C12 and C13
 in 3C 273.\label{fig1}}
\end{figure}

The components C11, C12 and C13 in \citet{kri01}, which have enough good quality data,
are chosen for the model fitting.
The location of a component at the first epoch is set as the
origin of the coordinate system, with $t_{\rm ob}=0$ and $\mu_{\rm ob}=0$.
The original proper motion
data and their fitted lines are shown in Figure \ref{fig1}. The
corresponding parameters are listed in Table \ref{tab1}. Figure \ref{fig2} shows
the predicted evolution of the Doppler factors $D$ of the components.

Model fitting results show that the observed proper motion
curves are sensitive to the viewing angle $\theta$ and to the 
initial velocity $\beta_0$. These two estimated parameters
have relatively small errors. The acceleration factor $k$, however,
has a larger error. The reason is that when a component
accelerates to an ultra-relativistic velocity, its apparent
velocity will remain almost constant, with
$\beta_{\rm app}\approx\cot(\theta/2)$, thus lateral observed data
will not dramatically contribute to the estimate of $k$. The $\chi^2$
values are lower than 1. This may due to the fact that the errors of the
original proper motion data have been over-estimated.
Table \ref{tab1} and Figures \ref{fig1} and \ref{fig2}, show that
the Doppler factors of all the three components are decreasing in the
last epoch.

\section{Discussion}
\subsection{Viewing angle}
The viewing angle plays an important role in the unification for radio-loud AGN.
It is believed that all radio-loud AGNs have a similar physical structure:
a supermassive black hole (probably spinning) located at their centre,
whose gravitational potential energy is the ultimate origin of the
AGN luminosity; an accretion disk surrounding the black hole, with relativistic jets
emitted perpendicular to it; broad line and narrow line regions,
and a dusty torus located at a larger distance from the innermost nucleus.
The appearance of AGNs, however, depends so strongly on orientation that our current
classification schemes are dominated by random pointing directions instead of
more grounded physical properties. Thus, to understand the intrinsic mechanism of
the AGNs, beaming effects induced by the viewing angle should be removed \citep{urr95}.

Besides the  apparent velocity measured by VLBI, other observational properties are
needed to estimate the viewing angle. The apparent velocity
$\beta_{\rm app}=\beta\sin\theta/(1-\beta\cos\theta)$ depends both on
$\theta$ and $\beta$, so $\theta$ cannot be derived only from $\beta_{\rm app}$.
If both apparent velocity $\beta_{\rm app}$ and Doppler boosting factor $D$ are known, then
the viewing angle can be calculated \citep{ghi93}. $\beta_{\rm app}$ can be directly
observed by VLBI. Currently, there are three ways to estimate $D$.
The first one is based on synchrotron self-Compton (SSC) model \citep{mar87}.
The second one depends on equipartition brightness temperature
\citep{rea94}. The last one can be derived from the flux density variability of
blazars \citep{lah99}. Compared to the LASM model, all of these three models need
extra physical parameters which are probably difficult to assess. For example, SSC model
requires the flux at the turnover frequency, which is usually substituted by 
the observed VLBI frequency in most cases, with an additional source of errors. 
Furthermore, these three models just provide the average values of the Doppler factors  
of jets, while the LASM model can trace the evolution of the Doppler factor as 
well as of the apparent velocity of the individual components.

The LASM model provides a direct and reliable way  to
estimate the viewing angle of a relativistic jet, provided that there are
enough high angular resolution VLBI monitoring data.
As shown in Section \ref{sec22}, the shape of a proper motion curve is
controlled by the viewing angle and other two parameters. Through nonlinear
fitting, all the parameters could be evaluated.
Furthermore, the acceleration will also influence the observed flux densities of jet
components (see Equation \ref{doppler}). If flux density data  are incorporated into
model fitting, the accuracy of
the estimation of viewing angle should be improved.

The model fitting results of 3C 273 show that the jet components
C11 C12 and C13 have different viewing angles (See Table
\ref{tab1}). This is clear from the  VLBI images \citep{kri01},
where different jet components move out along projected
trajectories with different position angles. The maximum
difference of position angles can reach 40 degree.
The difference of position angles as well as viewing angles may
indicate that there probably exists a super-massive binary black
hole system \citep{sun97} where the orbital motion of the
relativistic jet emerging from one massive black hole causes the
variation of viewing angles \citep{rie00}. Another possible
explanation is precession \citep{gow82}, and the precession angle
and period can be used to estimate the masses of the black holes
\citep{rom00}.

\subsection{Doppler factor\label{sec42}}
\begin{figure}
\includegraphics[width=10cm,height=10cm]{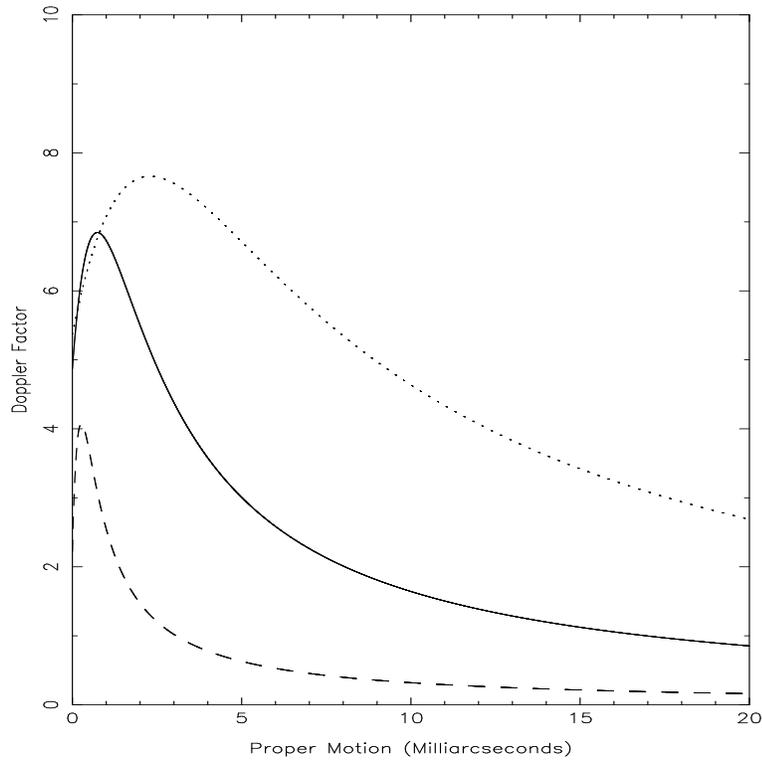}
\figcaption[f3.eps]{The evolution of the Doppler factors of the components
C11 (Dotted line), C12 (Dashed line) and C13 (Full line) in 3C 273. \label{fig2}}
\end{figure}

During the accelerating stage, the evolution of the Doppler factor
will affect the flux density variability of a jet component.
The observed flux density $S_{\rm ob}$ of a component is related to its intrinsic
flux density $S$ by $S_{\rm ob}=S D^{3+\alpha}$ \citep{bla79}, where $\alpha$
is the spectral index. A small variation of $D$ will cause a large change of
the observed flux density. As seen in Figure \ref{fig2}, during the
acceleration, the evolution of a Doppler factor is divided into two stages.
In the first stage, $D$ increases as the intrinsic speed of a component increases.
In the second stage, as the intrinsic speed keeps raising, $D$ will gradually 
decrease, going to values lower than 1.
Assuming $\alpha=0.5$, the observed flux densities of
C11, C12 and C13 are boosted up to the maximum level of 1267, 140 and 820 times 
respectively.
If we set $D=1$ as the detection threshold of VLBI, then the traceable lengths
of the components C11, C12 and C13 range roughly from 3 mas to 60 mas. This is in
agreement with the observations.

These considerations allow us to interpret the gaps between the inner and outer
jets of some radio loud AGNs, such as for instance 3C\,273 and M87.
For 3C\,273, the inner jet is about
100 mas long at 1.67GHz \citep{dav91}. The outer
jet is about $13''$ away from the inner jet and extends about $10''$
 \citep{sam01,mar01}. Inside this gap, no radiation is detected.
The possible reason is that continued acceleration of components will
cause Doppler deboosting. Therefore, at some distance, these components  will
disappear under the detection level of present observation.
However, the acceleration will end when the internal energy is mostly converted to
kinetic energy.  From then on, due to the interaction between the jets and the
medium of their host galaxy at sub-kpc scales, the components will decelerate and
and their Doppler factors will increase. Consequently,
the Doppler boosted components will appear again in the outer jet region.

\label{lastpage}

\clearpage

\clearpage

\clearpage
\begin{deluxetable}{llll}
\tabletypesize{\scriptsize}
\tablecaption{LASM model fitting
results of 3C 273 with confidence level of $68.3\%$.
DOF is the number of data used for fitting, $\theta$ is the viewing angle
of a component, $\beta_0$ is the initial velocity, and $k$ is the acceleration
factor. $\beta_{\rm app}^{0}$ and $\beta_{\rm app}^{f}$, $\gamma^{0}$ and $\gamma^{f}$, 
$D^{0}$ and $D^{f}$ denote the initial and final (at last observation epoch)
apparent velocities, Lorentz factors and Doppler boosting factors respectively. $D^{max}$ is the maximum Doppler
factor where $\beta=\cos\theta$ and $\gamma = D$. \label{tab1}}
\tablewidth{0pt}
\tablehead{Parameters & C11 & C12 & C13 }
\startdata
DOF & 26 & 26 & 29 \\
$\theta$ (degree) & 7.5$\pm$0.3 & 14.2$\pm$0.2 & 8.4$\pm$0.2 \\
$\beta_0$ & 0.954$\pm$0.009 & 0.7$\pm$0.2 & 0.937$\pm$0.015 \\
$k$ & 0.009$\pm$0.003 & 0.2$\pm$0.1  & 0.03$\pm$0.01 \\
$\beta_{\rm app}^{0}$ & 2.29 & 0.53 & 1.88 \\
$\beta_{\rm app}^{f}$ & 11.2 & 8.0 & 12.2 \\
$\gamma^{0}$ & 3.2 & 1.4 & 2.9 \\
$\gamma^{f}$ & 12.7 & 39.6  & 20.3 \\
$D^{0}$   & 5.4 & 2.2  & 4.8  \\
$D^{f}$   & 6.8 & 0.82  & 4.1  \\
$D^{max}$ & 7.7 & 4.1 & 6.8  \\
$\chi^2$ & 0.71  & 0.34  & 0.78 \\
\enddata
\end{deluxetable}

\end{document}